\documentclass[english,12pt]{article}
\usepackage{array}
\usepackage{graphicx}
\usepackage{amssymb}
\usepackage{amsmath}
\usepackage{multirow}
\usepackage{prettyref}
\usepackage{babel}
\usepackage{slashed}
\usepackage{units}
\usepackage[latin1]{inputenc}
\usepackage{amsfonts}
\usepackage{amssymb}
\usepackage{babel}
\usepackage{cite}
\def\@fmsl@sh#1#2#3{\m@th\ooalign{$\hfil#1\mkern#2/\hfil$\crcr$#1#3$}}
 \def\eq#1\en{\begin{equation}#1\end{equation}}
\def\s[#1,#2]{[#1\stackrel{\star}{,}#2]}
\def\sx[#1,#2]{[#1\stackrel{\star_{x}}{,}#2]}

\newcommand{\nc}{\newcommand}
\nc{\beq}{\begin{equation}}
\nc{\eeq}{\end{equation}}
\nc{\beqa}{\begin{eqnarray}}
\nc{\eeqa}{\end{eqnarray}}

\def\bc{\begin{center}}
\def\ec{\end{center}}

\def\to{\rightarrow}

\def\gsim{\mathrel{\mathpalette\atversim>}}

\def\bc{\begin{center}}
\def\ec{\end{center}}

\def\gsim{\mathrel{\rlap{\lower4pt\hbox{\hskip1pt$\sim$}}

    \raise1pt\hbox{$>$}}}       

\def\gsim{\mathrel{\rlap{\lower4pt\hbox{\hskip1pt$\sim$}}
    \raise1pt\hbox{$>$}}}       

\begin{document}
\makeatletter
\def\fmslash{\@ifnextchar[{\fmsl@sh}{\fmsl@sh[0mu]}}
\def\fmsl@sh[#1]#2{%
  \mathchoice
    {\@fmsl@sh\displaystyle{#1}{#2}}%
    {\@fmsl@sh\textstyle{#1}{#2}}%
    {\@fmsl@sh\scriptstyle{#1}{#2}}%
    {\@fmsl@sh\scriptscriptstyle{#1}{#2}}}
\def\@fmsl@sh#1#2#3{\m@th\ooalign{$\hfil#1\mkern#2/\hfil$\crcr$#1#3$}}
\makeatother

\thispagestyle{empty}
\begin{titlepage}
\boldmath
\begin{center}
  \Large {\bf What is modified gravity and how to differentiate it from particle dark matter?}
    \end{center}
\unboldmath
\vspace{0.2cm}
\begin{center}
{  {\large Xavier Calmet}\footnote{x.calmet@sussex.ac.uk}
and {\large Iber\^ e Kuntz}\footnote{ibere.kuntz@sussex.ac.uk}}
 \end{center}
\begin{center}
{\sl Physics $\&$ Astronomy, 
University of Sussex, Falmer, Brighton, BN1 9QH, United Kingdom 
}
\end{center}
\vspace{5cm}
\begin{abstract}
\noindent
An obvious criterion to classify theories of modified gravity is to identify their gravitational degrees of freedom and their coupling to the metric and the matter sector. Using this simple idea, we show that any theory which depends on the curvature invariants is equivalent to general relativity in the presence of new fields that are gravitationally coupled to the energy-momentum tensor. We show that they can be shifted into a new energy-momentum tensor. There is no a priori reason to identify these new fields as gravitational degrees of freedom or matter fields. This leads to an equivalence between dark matter particles gravitationally coupled to the standard model fields and modified gravity theories designed to account for the dark matter phenomenon. Due to this ambiguity, it is impossible to differentiate experimentally between these theories and any attempt of doing so should be classified as a mere interpretation of the same phenomenon.
\end{abstract}  
\end{titlepage}



\newpage
\section{Introduction}
\label{sec:intro}
General relativity and the standard model of particle physics have both been extremely successful in describing our universe both on cosmological scales as well as on microscopic scales. Despite this amazing success, some observations cannot be explained within these otherwise extremely successful models. For example, the cosmic microwave background, the rotation curves of galaxies or the bullet cluster to quote a few  \cite{Hooper:2009zm}, suggest that there is a new form of matter that does not shine in the electromagnetic spectrum. Dark matter is not accounted for by either general relativity or the standard model of particle physics \footnote{One should note though that the possibility Planck mass quantum black holes remnants \cite{Chen:2014jwq,Calmet:2014uaa} is not excluded, but it is difficult to find an inflationary model that produces them at the end of inflation}. While a large fraction of the high energy community is convinced that dark matter should be described by yet undiscovered new particles, it remains an open question whether this phenomenon requires a modification of the standard model or of general relativity.  Here we want to raise a slightly different question namely whether the distinction between modified gravity or new particles is always clear. We will show that this is not always the case.

Models of modified gravity are attractive given the frustrating success of the standard model at surviving to its confrontation with the data of the Large Hadron Collider. Modified theories of gravity have been developed in the hope of finding solutions to the dark matter or dark energy questions. All sorts of theories has been proposed in order to address these problems. Among them we can find higher derivative gravity theories (e.g. $f(R)$), the scalar-tensor theories (e.g. Brans-Dicke), the non-metric theories (e.g. Einstein-Cartan theory), just to cite a few, see \cite{Clifton:2011jh} for a substantial review.

In the context of quantum field theories, fields are just dummy variables as the action is formulated as a path integral over all field configurations. This implies a reparametrization invariance of field theories. In gravitational theories (see e.g. \cite{Calmet:2012eq}), this corresponds simply to the freedom to pick a specific frame to define one's model. The reparametrization invariance makes it difficult to differentiate between the plethora of models as depending on which field variables are picked, the very same model could appear to be very different in two different frames. One of the aims of this article is to apply a very simple and obvious criterion to classify gravitational theories. The idea is to identify their  gravitational degrees of freedom by looking at the poles in the field equations and carefully identifying the coupling of this poles to the metric and the energy-momentum tensor (matter sector). This enables one to unambiguously compare two gravitational models. Some work in this direction was done in the past \cite{Magnano:1995pv}, but the focus was given to the different action principles, namely the metric, metric-affine and affine formalisms. Here we present a broader approach which can be applied to any kind of theory independently of its action principle.

In this paper, we aim to propose a general framework where gravitational theories can be compared to each other so that we are able to classify them into different classes of physically equivalent theories. The classification method will be presented in Section \ref{sec:class} together with some examples. In Section \ref{sec:dark} we apply these ideas to the dark matter problem and show that the distinction between modified gravity or dark matter as a new particle is not always so clear. In particular, we show that any theory which depends on the curvature invariants is equivalent to general relativity in the presence of new fields that are gravitationally coupled to the energy-momentum tensor. We show that they can be shifted into a new energy-momentum tensor. Modified dark matter is thus equivalent to new degrees of freedom (i.e. particles) that are coupled gravitationally to regular matter. We then make the conclusions in Section \ref{sec:conc}.

\section{Classification of extended theories of gravity}
\label{sec:class}
Fields in a quantum field theory are dummy variables. The same applies to the metric in a gravitational theory. Therefore two apparently very different gravitational theories can actually turn out to be mathematically equivalent when expressed in the correct variables. A famous example is the $f(R)$ theory:
\begin{eqnarray}
S = \int d^4x \sqrt{-g} \left (\frac{1}{16 \pi G}f(R)+ \mathcal{L}_M\right)
\end{eqnarray}
where $f(R)$ is a polynomial of the Ricci scalar. When mapping the theory from the Jordan to the Einstein frame it becomes obvious that $f(R)$ is equivalent to usual general relativity with a scalar field that is gravitationally coupled to matter. Indeed, it is well known that after a Legendre transformation followed by a conformal rescaling $\tilde{g}_{\mu\nu} = f'(R) g_{\mu\nu}$, $f(R)$ theory can be put in the form \cite{DeFelice:2010aj}
\begin{eqnarray}
S &=& \int d^4x \sqrt{-\tilde g} \left(\frac{1}{16 \pi G}\tilde{R} - \frac{1}{2}\tilde{g}^{\mu\nu}\partial_\mu\phi\partial_\nu\phi - V(\phi)\right)\nonumber\\
&&+ \int d^4x \sqrt{-\tilde g} F^{-2}(\phi)\mathcal{L}_M(F^{-1}(\phi)\tilde g_{\mu\nu},\psi_M),
\label{eq:conf}
\end{eqnarray}
where
\begin{eqnarray}
\phi\equiv \sqrt{\frac{3}{16\pi G}}\log F,\\
F(\phi)\equiv f'(R(\phi)).
\end{eqnarray}
Hence all the matter fields acquires a universal coupling to a new scalar field $\phi$ through the factor $F^{-1}(\phi)$. Gauge bosons are exceptions since their Lagrangians are invariant under the metric rescaling. This simple example demonstrates that, despite the apparent simplicity of $f(R)$ which naively seems to only depends on the metric $g_{\mu\nu}$, the theory also contains an extra scalar degree of freedom.

This well known example can be generalized to any gravitational theory. A general gravitational theory, assuming that it is a metric theory, will have at least one metric tensor (if it is to have general relativity in some limit) and fields of different spins. We will assume that this theory can be described by an action $S=S[\phi^1_{\alpha_1},\ldots,\phi^n_{\alpha_n}]$, where $\phi^i_{\alpha_i}$ are the fields and $\alpha_i$ represents generically the number of indices, i.e. the type of the field (e.g. scalar, tensor, etc). The coupling of the gravitational degrees of freedom to matter ${\cal L_M}$ needs to be specified.  An algorithm to classify gravitational theories, in the sense of comparing two gravitational theories, can be designed as follows. 
\begin{itemize}
\item[1)] The first step then is to find all of the gravitational degrees of freedom of each theory.
\item[2)] Verify how these degrees of freedom couple to the metric tensor, to the matter degrees of freedom as well as to themselves.
\end{itemize}

The first step might sound obvious if what we have in mind are theories with a canonical Lagrangian. However, this is not the case for gravitational theories where degrees of freedom are hidden in terms in the action with higher number of derivatives (higher than two) acting on the metric as we have seen in the previous example. The identification of the degrees of freedom can be done as usual by linearizing the equations of motion around a fixed background $g_{\mu\nu}=g_{\mu\nu}^{(0)}+h_{\mu\nu}$, identifying the full propagator $\mathcal{P}_{\alpha\beta\mu\nu}$:
\begin{eqnarray}
\mathcal{D}_{\alpha\beta\mu\nu}h^{\mu\nu} = T_{\alpha\beta}\implies \mathcal{P}_{\alpha\beta\mu\nu} = \mathcal{D}_{\alpha\beta\mu\nu}^{-1}.
\end{eqnarray}
The position of the poles will reveal the different degrees of freedom hidden in a potential clumsy choice of variables. These degrees of freedom can be made explicit in the action, in some cases after the kinetic terms have been canonically normalized.

Having identified the degrees of freedom of the theories, we are left with the task of classifying their dynamics. For this purpose, there are two different approaches: one can either apply suitable transformations on the fields on the level of the Lagrangian in order to try to map one theory to another or one can proceed by calculating straightforwardly the equations of motion of each of them and then checking if they match in the end. It has to be stressed that both approaches lead to the same outcome and therefore we can conveniently choose how to proceed accordingly to the theory in hand.

In our previous example, we have shown that equation \eqref{eq:conf} implies that $f(R)$ theories can be described by a scalar field minimally coupled to general relativity. This means that $f(R)$ is formally equivalent to general relativity in the presence of a scalar field. Indeed, both theories have same degrees of freedom and their actions can be mapped into each other by field redefinitions. As can be seen from \eqref{eq:conf}, it is just a matter of choice whether the new scalar field $\phi$ belongs to the gravity sector or to the matter sector.

The same reasoning can be used for more general theories where it is also possible to identify new degrees of freedom besides the metric and the scalar of Equation \eqref{eq:conf}. In fact, an additional massive spin-2 is present in the generic theory $f(R,R_{\mu\nu}R^{\mu\nu},R_{\mu\nu\rho\sigma} R^{\mu\nu\rho\sigma})$ \cite{Magnano:2002uq,Nunez:2004ts,Chiba:2005nz}. As this is an important example for our considerations, we will now reproduce this well known fact using the results of \cite{Hindawi:1995an}. Consider the theory
\begin{eqnarray}
\label{eq:ac1}
S &=& \frac{1}{2\kappa^2}\int\mathrm{d}^4x\sqrt{-g}\left( R +\alpha R^2 + \beta R_{\mu\nu}R^{\mu\nu}+\gamma R_{\lambda\mu\nu\rho}R^{\lambda\mu\nu\rho}+\mathcal{L}_M(g_{\mu\nu},\phi_\alpha)\right),\\ \nonumber 
&=& \frac{1}{2\kappa^2} \int{ d^4x \sqrt{-g} 
        \left[ R + \frac{1}{6{m_0}^2} R^2 - \frac{1}{2{m_2}^2} 
            C^2 +\mathcal{L}_M(g_{\mu\nu},\phi_\alpha)\right] },\notag
\end{eqnarray}
where $C_{\mu\nu\rho\sigma}$ is the Weyl tensor, $m_0^{-2}=6\alpha+2\beta+2\gamma$ and $m_2^{-2}=-\beta-4\gamma$. The matter sector is represented by $\mathcal{L}_M(g_{\mu\nu},\phi_\alpha)$, where $\phi_\alpha$ denotes a set of arbitrary fields of any spin, but for the sake of the argument we will ignore the matter lagrangian for a while. Now we introduce a auxiliary scalar field $\lambda$:
\begin{eqnarray}
S &=& \frac{1}{2\kappa^2} \int{ d^4x \sqrt{-g} \left[
       R+ \frac{1}{6{m_0}^2} R^2 - \frac{1}{6{m_0}^2} (R-3m_0^2\lambda)^2 - \frac{1}{2{m_2}^2} 
            C^2
       \right] }\\ \nonumber 
&=& \frac{1}{2\kappa^2} \int{ d^4x \sqrt{-g} \left[
       (1+\lambda)R - \frac32 m_0^2\lambda^2 - \frac{1}{2{m_2}^2} 
            C^2
       \right] }\\ \nonumber 
&=& \frac{1}{2\kappa^2} \int{ d^4x \sqrt{-g} \left[
       e^\chi R - \frac32 m_0^2(e^\chi-1)^2 - \frac{1}{2{m_2}^2} 
            C^2
       \right] }.
\end{eqnarray}
In the last line, we made the redefinition $\chi = \log(1+\lambda)$. The equation of motion for $\lambda$ is algebraic and given by $R=3m^2\lambda$. Substituting this back into the action gives the original theory back. Therefore, both theories are equivalent. Now we can perform a conformal transformation $\tilde g_{\mu\nu} = e^\chi g_{\mu\nu}$
\begin{eqnarray}
S= \frac{1}{2\kappa^2} \int{ d^4x \sqrt{-\tilde g} \left[ \tilde R 
           - \tfrac32 \left(\tilde \nabla\chi\right)^2 
           - \tfrac32 {m_0}^2\left(1-e^{-\chi}\right)^2
           - \frac{1}{2{m_2}^2} \tilde C^2
           \right] },
\end{eqnarray}
where we have used the fact that $C^2$ is invariant under conformal transformations. Now we can rewrite the above action as
\begin{eqnarray}
S     &=& \frac{1}{2\kappa^2} \int{ d^4x \sqrt{-\tilde g} \left[ \tilde R 
           - \tfrac32 \left(\tilde \nabla\chi\right)^2 
           - \tfrac32 {m_0}^2\left(1-e^{-\chi}\right)^2 - \frac1{2m^2_2} \left( \tilde R_{\lambda\mu\nu\rho}\tilde R^{\lambda\mu\nu\rho} 
               - 2\tilde R_{\mu\nu}\tilde R^{\mu\nu} + \tfrac13 \tilde R^2 \right)
           \right] } \nonumber    \\ \nonumber 
     &=& \frac{1}{2\kappa^2} \int d^4x \sqrt{-\tilde g} \bigg[ \tilde R 
           - \tfrac32 \left(\tilde \nabla\chi\right)^2 
           - \tfrac32 {m_0}^2\left(1-e^{-\chi}\right)^2 - \frac{1}{{m_2}^2} 
            \left( \tilde R_{\mu\nu}\tilde R^{\mu\nu} - \tfrac{1}{3}\tilde R^2 \right)\notag \\
&&- \frac1{2m^2_2} \left( \tilde R_{\lambda\mu\nu\rho}\tilde R^{\lambda\mu\nu\rho} 
               - 4\tilde R_{\mu\nu}\tilde R^{\mu\nu} + \tilde R^2\right)\bigg]. 
               \end{eqnarray}
Due to the Gauss-Bonnet theorem, the last term of the last line vanishes and we end up with
\begin{equation}
\label{eq:ac2}
S = \frac{1}{2\kappa^2} \int d^4x \sqrt{-\tilde g} \left[ \tilde R 
           - \tfrac32 \left(\tilde \nabla\chi\right)^2 
           - \tfrac32 {m_0}^2\left(1-e^{-\chi}\right)^2 - \frac{1}{{m_2}^2} 
            \left( \tilde R_{\mu\nu}\tilde R^{\mu\nu} - \tfrac{1}{3}\tilde R^2 \right)\right].
\end{equation}
We then add a auxiliary symmetric tensor field $\tilde\pi_{\mu\nu}$:
\begin{equation}
\label{eq:ac3}
S = \frac{1}{2\kappa^2} \int{ d^4x \sqrt{-\tilde g} \left[ \tilde R 
           - \tfrac32 \left(\tilde\nabla\chi\right)^2 
           - \tfrac32 {m_0}^2\left(1-e^{-\chi}\right)^2
           - \tilde G_{\mu\nu}\tilde \pi^{\mu\nu} 
           + \tfrac14 {m_2}^2 \left( \tilde\pi_{\mu\nu}\tilde\pi^{\mu\nu} - \tilde\pi^2 
\right)
           \right]}.
\end{equation}
where $\tilde\pi = \tilde\pi_{\mu\nu}\tilde G^{\mu\nu}$ and $\tilde G_{\mu\nu}$ is the Einstein tensor in the Einstein frame. The $\tilde\pi_{\mu\nu}$ equation of motion is
\begin{eqnarray}
\tilde G_{\mu\nu} = \tfrac{1}{2}m^2_2 \left( \tilde\pi_{\mu\nu} - \tilde g_{\mu\nu}\tilde\pi \right),
\end{eqnarray}
which can be written in the form
\begin{eqnarray}
\tilde\pi_{\mu\nu}=2{m_2}^{-2}\left(\tilde R_{\mu\nu}-\frac{1}{6}
\tilde g_{\mu\nu}\tilde R\right).
\end{eqnarray}
Substituting this equation of motion back into the action \eqref{eq:ac3} leads to the action \eqref{eq:ac2}, thus they are equivalent. Therefore, we have proven the equivalence between the actions \eqref{eq:ac1} and \eqref{eq:ac3}. From action \eqref{eq:ac3}, we can see that our original theory is equivalent to general relativity in the presence of a canonical scalar field and a non-canonical symmetric rank-2 tensor field. It is tempting to say that $\tilde\pi_{\mu\nu}$ is a spin-2 field, but this is not obvious at this stage. So far, $\tilde\pi_{\mu\nu}$ describes 10 degrees of freedom, while a massive spin-2 describes only 5. In the simplest case of a free spin-2 field $\phi_{\mu\nu}$ on a flat spacetime, such field is described by the Pauli-Fierz action. The divergence and the trace of its equation of motion imply the conditions:
\begin{eqnarray}
\partial^\mu\phi_{\mu\nu}=0,\quad \phi=0,
\end{eqnarray}
which constrains the number of degrees of freedom to 5. For a general spin-2 field though, the above conditions are no longer satisfied, but we can still find generalized conditions in order to reduce the number of degrees of freedom to 5. From the trace of the $\tilde g_{\mu\nu}$ equation of motion and from the divergence of the $\tilde\pi_{\mu\nu}$ equation of motion we find:
\begin{eqnarray}
&\tilde\nabla^{\mu} \left( \tilde\pi_{\mu\nu} - g_{\mu\nu}\tilde\pi \right) = 0, \\
 &  \tilde\pi - {m_2}^{-2} \left[ \left(\tilde\nabla\chi\right)^2 
            + 2{m_0}^2 \left(1-e^{-\chi}\right)^2 \right] = 0.
\end{eqnarray}
The above conditions give 5 constraints, thus reduces the number of degrees of freedom described by $\tilde\pi_{\mu\nu}$ to 5. Now $\tilde\pi_{\mu\nu}$ is a pure spin-2 field. Furthermore, if we linearize our theory, the above conditions give Pauli-Fierz conditions back and, therefore, $\tilde\pi_{\mu\nu}$ would produce a canonical spin-2 field. Thus, we managed to find a spin-2 field, even though it does not appear canonically in the Lagrangian.

To canonically normalize the field $\tilde\pi_{\mu\nu}$, we need to perform another transformation on the metric. We start by writing the Lagrangian \eqref{eq:ac3} in the form
\begin{eqnarray}
S &=& \frac{1}{2\kappa^2} \int d^4x \sqrt{-\tilde g} \bigg\{\bigg[\big(1+\tfrac12\tilde\pi\big)\tilde g^{\mu\nu}-\tilde \pi^{\mu\nu}\bigg]\tilde R_{\mu\nu}
  + \tfrac14 {m_2}^2 \bigg( \tilde\pi_{\mu\nu}\tilde\pi^{\mu\nu} - \tilde\pi^2\bigg)\\ \nonumber
  &&- \tfrac32 \left(\tilde\nabla\chi\right)^2 - \tfrac32 {m_0}^2\bigg(1-e^{-\chi}\bigg)^2 \bigg]\bigg\}.
\end{eqnarray}
To get a canonical Einstein-Hilbert term, we need to redefine the metric as
\begin{eqnarray}
\sqrt{-\bar g}\bar{g}^{\mu\nu} = \sqrt{-\tilde g}\left[\left(1+\tfrac12\tilde\pi\right)\tilde g^{\mu\nu}-\tilde\pi^{\mu\nu}\right],
\end{eqnarray}
which leads to the transformations
\begin{eqnarray}
\label{eq:trans}
\bar g^{\mu\nu} &=& (\det A)^{-1/2}\tilde g^{\mu\lambda}A_\lambda^\nu\\
A_\lambda^\nu &=& (1+\tfrac12 \phi)\delta_\lambda^\nu-\phi_\lambda^\nu.
\end{eqnarray}
We have introduced the new notation $\phi_\mu^\nu=\tilde\pi_\mu^\nu$ to emphasize that the indices of $\phi_{\mu\nu}$ are raised and lowered using $\bar g_{\mu\nu}$, while the indices of $\tilde\pi_{\mu\nu}$ were raised and lowered using $\tilde g_{\mu\nu}$. Therefore, in the new variables the Lagrangian reads
\begin{eqnarray}
   S &=& \frac{1}{2\kappa^2} \int d^4x \sqrt{-\bar g} \bigg[ \bar R
           - \tfrac{3}{2}\left(A^{-1}(\phi_{\sigma\tau})\right)_{\mu}^{\ \ \nu}
                  \bar \nabla^\mu\chi\bar\nabla_\nu\chi
           - \tfrac{3}{2}\left(\textrm{det}A(\phi_{\sigma\tau})\right)^{-1/2}
                  \left(1-e^{-\chi}\right)^2
              \\  \nonumber
           &&- \bar g^{\mu\nu} \left( C^\lambda_{\ \ \mu\rho}(\phi_{\sigma\tau}) 
                  C^\rho_{\ \ \nu\lambda}(\phi_{\sigma\tau})
              - C^\lambda_{\ \ \mu\nu}(\phi_{\sigma\tau}) 
                  C^\rho_{\ \ \rho\lambda}(\phi_{\sigma\tau}) \right)
              \\ \nonumber
           &&+ \tfrac{1}{4}{m_2}^2 
              \left(\textrm{det}A(\phi_{\sigma\tau})\right)^{-1/2}
              \left( \phi_{\mu\nu}\phi^{\mu\nu} - \phi^2 \right)
           \bigg],
\end{eqnarray}
where
\begin{eqnarray}
C^\lambda_{\ \ \mu\nu} = \tfrac12 (\tilde g^{-1})^{\lambda\rho}(\bar\nabla_\mu\tilde g_{\nu\rho} + \bar\nabla_\nu\tilde g_{\mu\rho} - \bar\nabla_\rho\tilde g_{\mu\nu}).
\end{eqnarray}
Due to the transformation \eqref{eq:trans}, the metric $\tilde g=\tilde g(\phi_{\mu\nu})$ now depends on the spin-2 field. Thus the spin-2 kinetic term appears explicitly in the action through $C^\lambda_{\ \ \mu\nu}$.

In the presence of external matter the argument goes in the same way, except that after performing the transformations the matter Lagrangian becomes $\mathcal{L}_M(e^{-\chi}\tilde g_{\mu\nu}(\phi_{\sigma\tau}),\phi_\alpha)$ and the action reads
\begin{eqnarray}
   S &=& \frac{1}{2\kappa^2} \int d^4x \sqrt{-\bar g} \bigg[ \bar R
           - \tfrac{3}{2}\left(A^{-1}(\phi_{\sigma\tau})\right)_{\mu}^{\ \ \nu}
                  \bar \nabla^\mu\chi\bar\nabla_\nu\chi
           - \tfrac{3}{2}\left(\textrm{det}A(\phi_{\sigma\tau})\right)^{-1/2}
                  \left(1-e^{-\chi}\right)^2
              \\ \nonumber
           &&- \bar g^{\mu\nu} \left( C^\lambda_{\ \ \mu\rho}(\phi_{\sigma\tau}) 
                  C^\rho_{\ \ \nu\lambda}(\phi_{\sigma\tau})
              - C^\lambda_{\ \ \mu\nu}(\phi_{\sigma\tau}) 
                  C^\rho_{\ \ \rho\lambda}(\phi_{\sigma\tau}) \right)
              \\  \nonumber
           &&+ \tfrac{1}{4}{m_2}^2 
              \left(\textrm{det}A(\phi_{\sigma\tau})\right)^{-1/2}
              \left( \phi_{\mu\nu}\phi^{\mu\nu} - \phi^2 \right)
	   +\bar{\mathcal{L}}_M(e^{-\chi}\tilde g_{\mu\nu}(\phi_{\sigma\tau}),\phi_\alpha)\bigg].
\end{eqnarray}
where 
\begin{eqnarray}
\bar{\mathcal{L}}_M = e^{-2\chi}(\det{A}(\phi_{\mu\nu}))^{-1/2}\mathcal{L}_M.
\end{eqnarray}
We see that, in general, external matter couples minimally to the usual graviton through the Jacobian $\sqrt{-\bar g}$ and non-minimally to the fields $\chi$ and $\phi_{\mu\nu}$.

In the following, we will calculate explicitly the coupling between external matter and the additional degrees of freedom $\chi$ and $\phi_{\mu\nu}$. Consider a matter Lagrangian being composed of a scalar, a vector and a spinor field:
\begin{eqnarray}
\mathcal{L}_M = \mathcal{L}_0 + \mathcal{L}_1 + \mathcal{L}_{1/2},
\end{eqnarray}
where
\begin{eqnarray}
\mathcal{L}_0 &=& \tfrac12\nabla_\mu\sigma\nabla^\mu\sigma\\
\mathcal{L}_1 &=& -\tfrac14 F_{\mu\nu}F^{\mu\nu}\\
\mathcal{L}_{1/2} &=& i\bar\psi\slashed\partial\psi.
\end{eqnarray}
After transforming the metric to $\bar g_{\mu\nu}$ (i.e., $g_{\mu\nu}\to\tilde g_{\mu\nu}\to\bar g_{\mu\nu}$), we get
\begin{eqnarray}
\bar{\mathcal{L}}_0 &=& \tfrac12 e^{-\chi} (A^{-1})_\alpha^{\ \ \nu}\bar g^{\alpha\mu}\nabla_\mu\sigma\nabla_\nu\sigma,\\
\bar{\mathcal{L}}_1 &=& -\tfrac{1}{4}(\det A)^{1/2}(A^{-1})_\rho^{\ \ \mu}(A^{-1})_\lambda^{\ \ \nu}\bar g^{\rho\alpha}\bar g^{\lambda\beta}F_{\mu\nu}F_{\alpha\beta},\\ 
\bar{\mathcal{L}}_{1/2} &=& e^{-\chi}(A^{-1})_\alpha^{\ \ \nu}i\bar\psi\bar g^{\alpha\mu}\gamma_\mu\partial_\nu\psi,
\end{eqnarray}
and $\bar{\mathcal{L}}_M = \bar{\mathcal{L}}_0 + \bar{\mathcal{L}}_1 + \bar{\mathcal{L}}_{1/2}$. One can also consider interaction terms, namely the Yukawa interaction and the gauge interactions for spinor-vector fields and scalar-vector fields and study how the are affected by the metric redefinition:
\begin{eqnarray}
\mathcal{L}_{\text{Yukawa}} &=& -g\bar\psi\phi\psi,\\
\mathcal{L}_0 &=& \tfrac12 (D_\mu\sigma)^\dagger (D^\mu\sigma) = \frac12 \nabla_\mu\sigma\nabla^\mu\sigma + e^2 A_\mu A^\mu\sigma^2,\\
\mathcal{L}_{1/2} &=& i\bar\psi\slashed D\psi = i\bar\psi\gamma^\mu \nabla_\mu\psi - eA_\mu\bar\psi\gamma^\mu\psi,
\end{eqnarray}
where $D_\mu = \nabla_\mu + ieA_\mu$. After transforming the metric to $\bar g_{\mu\nu}$ (i.e., $g_{\mu\nu}\to\tilde g_{\mu\nu}\to\bar g_{\mu\nu}$), one finds
\begin{eqnarray}
\bar{\mathcal{L}}_{\text{Yukawa}} &=& -e^{-2\chi}(\det A)^{-1/2}g\bar\psi\phi\psi,\\
\bar{\mathcal{L}}_0 &=& \tfrac12 e^{-\chi} (A^{-1})_\alpha^{\ \ \nu}\bar g^{\alpha\mu}(\nabla_\mu\sigma\nabla_\nu\sigma+e^2 A_\mu A_\nu\sigma^2),\\
\bar{\mathcal{L}}_{1/2} &=& e^{-\chi}(A^{-1})_\alpha^{\ \ \nu}\bar g^{\alpha\mu}(i\bar\psi\gamma_\mu\partial_\nu\psi-e A_\mu\bar\psi\gamma_\nu\psi),
\end{eqnarray}
and 
$\bar{\mathcal{L}}_M = \bar{\mathcal{L}}_0 + \bar{\mathcal{L}}_1 + \bar{\mathcal{L}}_{1/2}+\bar{\mathcal{L}}_{\text{Yukawa}}$.
We note that the massive spin-2 field couples to all matter fields of spin 0, 1/2 and 2 because of  the matrix $A$. On the other hand, the scalar field $\chi$ does not couple to spinors. The masses of the spin 0 and massive spin 2 gravitational fields can be tuned by adjusting the coefficients of the action, on the other hand their interactions with matter fields, while not always universal, is fixed by the gravitational coupling constant. As usual, the massless graviton couples universally and gravitationally to matter fields.

\section{Application to dark matter}
\label{sec:dark}

As already emphasized, astrophysical and cosmological evidence for dark matter is overwhelming.  Several explanations have been proposed to explain the dark matter phenomenon. These models are usually classified into two categories: modifications of Einstein's general relativity or modifications of the standard model in the form of new particles. The aim of this section is to point out that these two categories are not so different after all. In fact, every modified gravity model has new degrees of freedom besides the usual massless graviton.

The first attempt to explain galaxy rotation curves by a modification of Newtonian dynamics is due to Milgrom \cite{Milgrom:1983ca}. While Milgrom's original proposal was non-relativistic and very phenomenological, more refined theories have been proposed later on, including Bekenstein's TeVeS theory \cite{Bekenstein:2004ne}, Moffat's modified gravity (MOG) \cite{Moffat:2005si} and Mannheim's conformal gravity \cite{Mannheim:2011ds}, which are relativistic. While these theories seem to be able to explain the rotation curves of the galaxies (see e.g. \cite{Famaey:2011kh} for a recent MOND review where the observational successes are discussed in details), it is more difficult to imagine how they would explain the bullet cluster observations or the agreement of the CMB observation with the standard cosmological model $\Lambda$CMB which posits the existence of cold dark matter. We shall not dwell on the question of the viability of modified gravity as we may simply not yet have found the correct model. However, we merely point out that if such a theory exists, it will not be necessarily very different from a model involving particles as dark matter.

Indeed, whatever this realistic theory might be, it can be parameterized by a function $f(R,R_{\mu\nu},R_{\mu\nu\rho\sigma},\phi_\alpha)$ modelled using effective theory techniques. Here $R$ is the Ricci scalar, $R_{\mu\nu}$ is the Ricci tensor and $\phi_\alpha$ denotes collectively any type of field that is also responsible for the gravitational interaction. In terms of effective field theory, any theory of modified gravity can be described by
\begin{eqnarray}
S=\frac{1}{16 \pi G} \int d^4x \sqrt{-g} f(R,R_{\mu\nu},R_{\mu\nu\rho\sigma},\phi_\alpha) + \int d^4x \sqrt{-g} \mathcal{L}_M
\end{eqnarray}
where $G$ is Newton's constant. We are only assuming diffeomorphism invariance and the usual space-time and gauge symmetries for the matter content described by the Lagrangian $\mathcal{L}_M$. A successful model should lead to a modification of Newton's potential that fits, e.g., the galaxy rotation curves. It is not difficult to imagine that the standard Newtonian term $1/r$ would come from the usual massless spin-2 graviton exchange while the non-Newtonian terms would have to be generated by the new degrees of freedom. Clearly, it is not straightforward to come up with such a model, however, as mentioned before, there are a few known examples.

While it is obvious that new degrees of freedom are included when $\phi_\alpha$ is added to the function $f$ as in \cite{Moffat:2005si}, it is much less clear how they are identified when the theory is a function of the curvature invariants only as we stressed before. Hence we will restrict ourselves to the theory $f(R,R_{\mu\nu},R_{\mu\nu\rho\sigma} )$. From the arguments made at the end of Section \ref{sec:class}, we know that this theory is equivalent to general relativity in the presence of a scalar field and of a massive spin-2 field. Therefore, there is no difference between introducing new particles and introducing modifications on gravity, which raises the question of whether it is possible to differentiate experimentally between models of modified gravity and particle dark matter. Nonetheless, since the massive spin-2 particle is a ghost, this result also suggests that a good dark matter model is very likely to be described either by an $f(R)$ theory and hence a scalar field.

Any modification of gravity that has general coordinate invariance as a symmetry can be reformulated, using appropriate variables, as usual general relativity  accompanied by new degrees of freedom. We have seen that these new degrees of freedom may not couple universally to matter. Modified gravity can thus be seen as a model with new dark matter particles that are very weakly coupled to the standard model. These apparently very different models describe the same physics as their actions are related by simple variables transformations. This may provide a simple way to modified gravity proponents to explain bullet clusters experiments or the cosmic microwave background.

\section{Conclusions}
\label{sec:conc}
In this paper, we proposed a classification scheme for gravitational theories. In particular, we showed the equivalence between the broad class of theories $f(R,R_{\mu\nu},R_{\mu\nu\rho\sigma} )$ and general relativity in the presence of additional matter fields, namely a scalar and a massive spin-2 field. We have shown that these new degrees of freedom can be shifted into a redefined stress-energy tensor and that they will coupled gravitationally to the matter fields introduced in the model. We conclude that any attempt to modify the Einstein-Hilbert action, preserving the underlying symmetry, leads to new degrees of freedom, i.e., new particles. In that sense, this is not different from including new matter fields by hand in the matter sector that are coupled gravitationally to the standard model matter fields. Assuming that models of modified gravity preserve diffeomorphism invariance, we have shown that they are equivalent to general relativity with new degrees of freedom coupled gravitationally to the fields of the standard model. From that point of view, there is a duality between models of modified gravity and a particle physics models with new fields that are coupled gravitationally to the standard model. 

These results may make it easier to analyse the physics of models of dark matter involving a  modification of gravity and, in particular, the fact that they are dual to some very weakly coupled dark matter model could help to resolve the apparent conflict with bullet cluster observations. 

While we focussed on dark matter in this paper as an application for the classification of extended theories of gravity we proposed, another obvious application would be to the physics of gravitational waves for which extended theories of gravity are also important, see e.g. \cite{Capozziello:2011au,DeLaurentis:2016jfs,Capozziello:2015nga,Calmet:2016sba}.

\noindent{\it Acknowledgments:}
This work is supported in part by the Science and Technology Facilities Council (grant number  ST/L000504/1) and by the National Council for Scientific and Technological Development (CNPq - Brazil).


\bigskip{}

\end{document}